\documentclass[10pt, conference, compsocconf]{IEEEtran}
\usepackage{graphicx}
\usepackage{algorithm}
\usepackage{algpseudocode}
\begin{document}
\title{Using Machine Learning to reduce the energy wasted in Volunteer Computing Environments}
\author{
\IEEEauthorblockN{A. Stephen McGough and Matthew Forshaw}
\IEEEauthorblockA{School of Computing\\
Newcastle University\\
Newcastle, UK\\
Email: \{stephen.mcgough,matthew.forshaw\}\\@newcastle.ac.uk}
\and
\IEEEauthorblockN{John Brennan, Noura Al Moubayed, Stephen Bonner}
\IEEEauthorblockA{Department of Computer Science\\
Durham University\\
Durham, UK\\
Email: \{john.brennan, noura.al-moubayed, s.a.r.bonner\}\\@durham.ac.uk}
}
\IEEEoverridecommandlockouts
%
\maketitle
\begin{abstract}High Throughput Computing (HTC) provides a convenient mechanism for running thousands of tasks. Many HTC systems exploit computers which are provisioned for other purposes by utilising their idle time -- volunteer computing. This has great advantages as it gives access to vast quantities of computational power for little or no cost. The downside is that running tasks are sacrificed if the computer is needed for its primary use. Normally terminating the task which must be restarted on a different computer -- leading to wasted energy and an increase in task completion time. We demonstrate, through the use of simulation, how we can reduce this wasted energy by targeting tasks at computers less likely to be needed for primary use, predicting this idle time through machine learning. By combining two machine learning approaches, namely Random Forest and MultiLayer Perceptron, we save 51.4\% of the energy without significantly affecting the time to complete tasks.
\end{abstract}
\IEEEpeerreviewmaketitle
%
\section{Introduction}
Many research problems that we face today require the execution of large computational workloads which can seriously hinder progress. To mitigate the impact of these computational workloads two main approaches have become prevalent -- those of High Performance Computing (HPC) and High Throughput Computing (HTC). Both of these approaches take advantage of running the workload across many computational units\footnote{Here we shall refer to these as computers, without loss of generality.}. HPC is used in cases where the workload requires inter-communication between computers, whilst HTC allows the workload to be decomposed into separate, non interacting, tasks\footnote{In the literature these may be referred to as task or jobs. We will use the term tasks in this work without loss of generality.} which run independently.

Although HTC can be seen as a `simpler' problem than HPC -- removing the need to simultaneously provision large numbers of computers, handle inter-computer communication issues, and issues over communication versus computation -- the efficient deployment and execution of HTC tasks have their own problems. These include: the timely deployment of tasks; ensuring the, potentially thousands or more, tasks run to completion; and the deployment / collection of the required data and code. These problems are compounded when the tasks are run on computers which are not owned or provisioned for HTC tasks -- often referred to as volunteer computing -- as HTC tasks are sacrificed when the computer is required for its primary role. Despite this, volunteer computing is often desired due to the significant computational power it provides -- often at little or no cost to the user. Two common platforms for volunteer HTC are HTCondor~\cite{htcondor} and BOINC~\cite{boinc}.
\IEEEpubidadjcol

In volunteer computing HTC tasks are normally run when the computer is `idle' and not performing their primary role. However, if the computer is needed again for its primary role then the HTC system needs to relinquish the computer. This may be through the termination of the HTC task, task suspension or migrating the task to a different computer. Task termination can be performed in any environment and for any task, however, suspension and migration require support from both the underlying infrastructure and the task being performed. Thus, in many cases HTC users will default to task termination in which case the HTC system will attempt to re-run the task on a different computer. 

Re-running the task on a different computer leads to two detrimental impacts: an increase in the time, in excess of the task execution time, the user must wait to obtain results -- referred to as task overhead, and an increase in energy consumed due to the, potentially multiple, aborted task executions. Thus we seek to reduce the energy consumed, whilst at the same time maintaining or reducing overhead. 
Ideally we wish to identify those computers which are less likely to be required for their primary use during the execution time of the task. This will lead to a reduction in wasted energy as the task will not have aborted executions. However, this may increase the overhead due to delay in finding an appropriate computer. Alternatively it may reduce overhead as time will not be wasted on partial executions. 

In order to identify those computers less likely to be needed for their primary use we use machine learning to predict the time, for each computer, between primary usage -- referred to as idle time. We can then select the computer with the largest predicted idle time to run a task. We evaluate two machine learning approaches for predicting the lengths of idle periods. These are Random Forest~\cite{liaw2002classification} and MultiLayer Perceptron~\cite{ruck1990multilayer} a form of feedforward artificial neural network~\cite{DLBook}. We also combine these using various Ensemble techniques~\cite{ensemble} in order to identify the `best' prediction. In an ideal scenario we would also seek to predict the execution time of the tasks submitted to the HTC system, however, prior research has shown that this is not easy for the user~\cite{lee}. Nor is this easy for machine learning as although there is significant correlation between tasks submitted at the same time there is little correlation to future tasks.

In Section \ref{HTC-SIM} we discuss the HTC-Sim which we used in order to evaluate our different machine learning approaches. This is followed by a discussion of the dataset made available with the HTC-Sim system in Section \ref{data}. Related work is presented in Section \ref{related}. We present our machine learning approaches to predicting computer idle time in Section \ref{ml} before discussing the experimental setup in Section \ref{experiment}. We present the results and analysis in Section \ref{results} before providing conclusions and future directions in Section \ref{conc}.

\section{HTC-Sim System}
\label{HTC-SIM}

The HTC-Sim System~\cite{htc-sim-short} is a trace-driven simulation framework for a generic High Throughput Computing system. It is capable of simulating both dedicated computer resources and computers which are provided on a voluntary basis. Each run of HTC-Sim consumes a number of trace log files -- a trace file for the primary users (referred to as interactive users) of the volunteer computers and a trace file for the tasks submitted to the HTC system (referred to as tasks from High-Throughput users). Different scheduling algorithms can be developed and deployed within the simulation system and evaluated using the provided metrics.

\begin{figure}[!t]
\centering
\includegraphics[width=8cm]{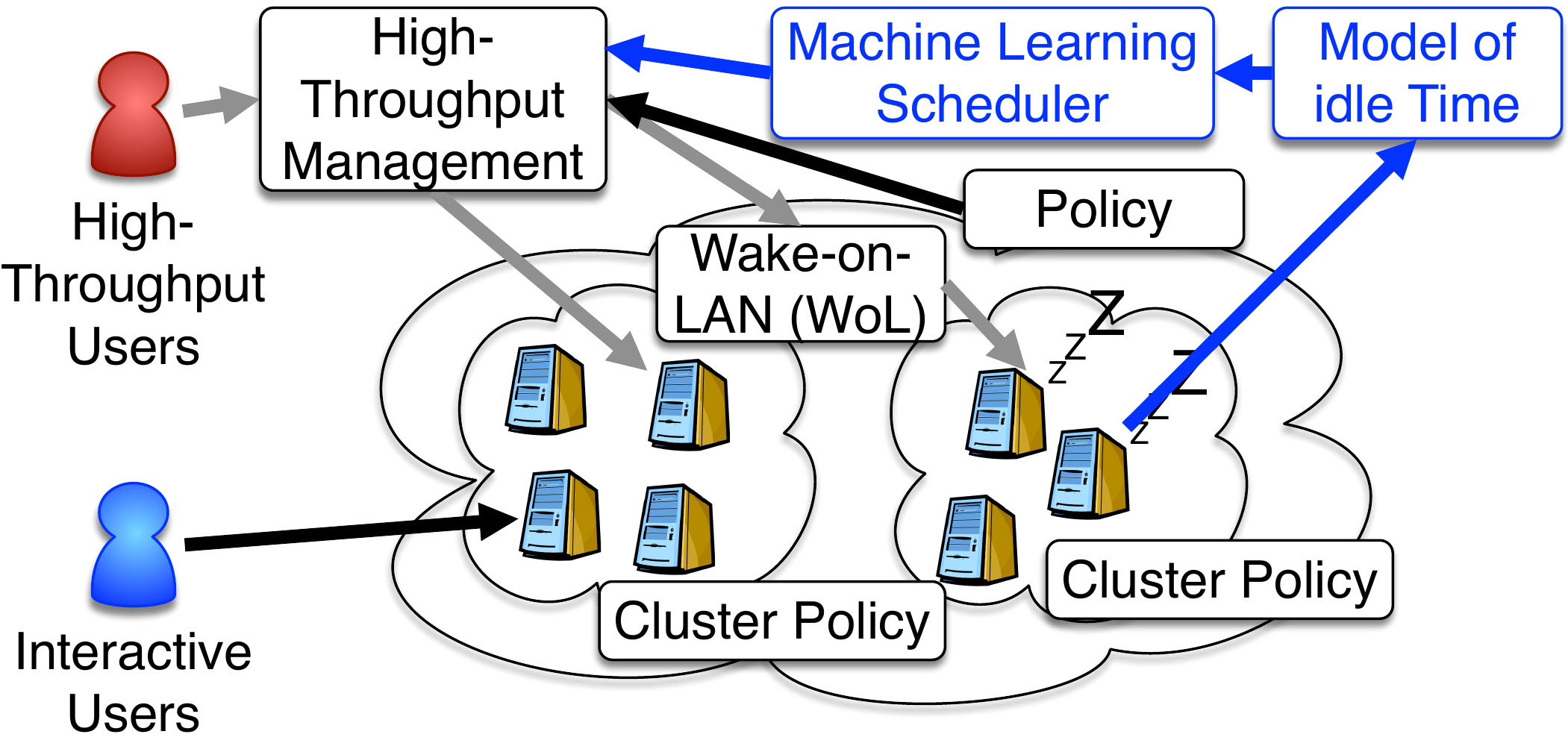}
\caption{The HTC-Sim model}
\label{fig:condor}
\end{figure}

Figure \ref{fig:condor} illustrates the model view of HTC-Sim. Computers within the system may be in one of three states: i) servicing the primary user of the computer, ii) executing a HTC task, or, iii) in an idle state. The idle state can be sub-divided into: a) idle and powered up, or, b) idle and in sleep state. Computers will transition from idle powered up to idle sleep after a pre-defined period of inactivity -- thus minimising wasted energy. This is one example of the Cluster Policy which covers such issues as when the computers should reboot (for software updates), whether HTC tasks can be performed on the computer and the minimum time between a primary user logging out and a HTC task being deployed. The current state of the HTC system is maintained by the High-Throughput Management which handles task deployment along with the transferring of files. If a computer is in idle sleep, but is required for a task the High-Throughput Management can wake up the computer, provided that this is currently allowed by policy.

We have augmented Figure \ref{fig:condor} with a machine learning scheduler and a service which constructs models of individual computer idle times -- highlighted in blue.

\section{Exemplar Data}
\label{data}
The HTC-Sim system is provided with a set of exemplar trace-logs for a university environment. Containing 1386 computers separated into 37 clusters. Where computers within a cluster are assumed to be identical hardware whilst computers in different clusters may be different. Each computer type is modelled with an energy consumption rate for the states of: active, idle and sleep. 

\begin{figure}[!b]
\centering
\includegraphics[width=8.5cm]{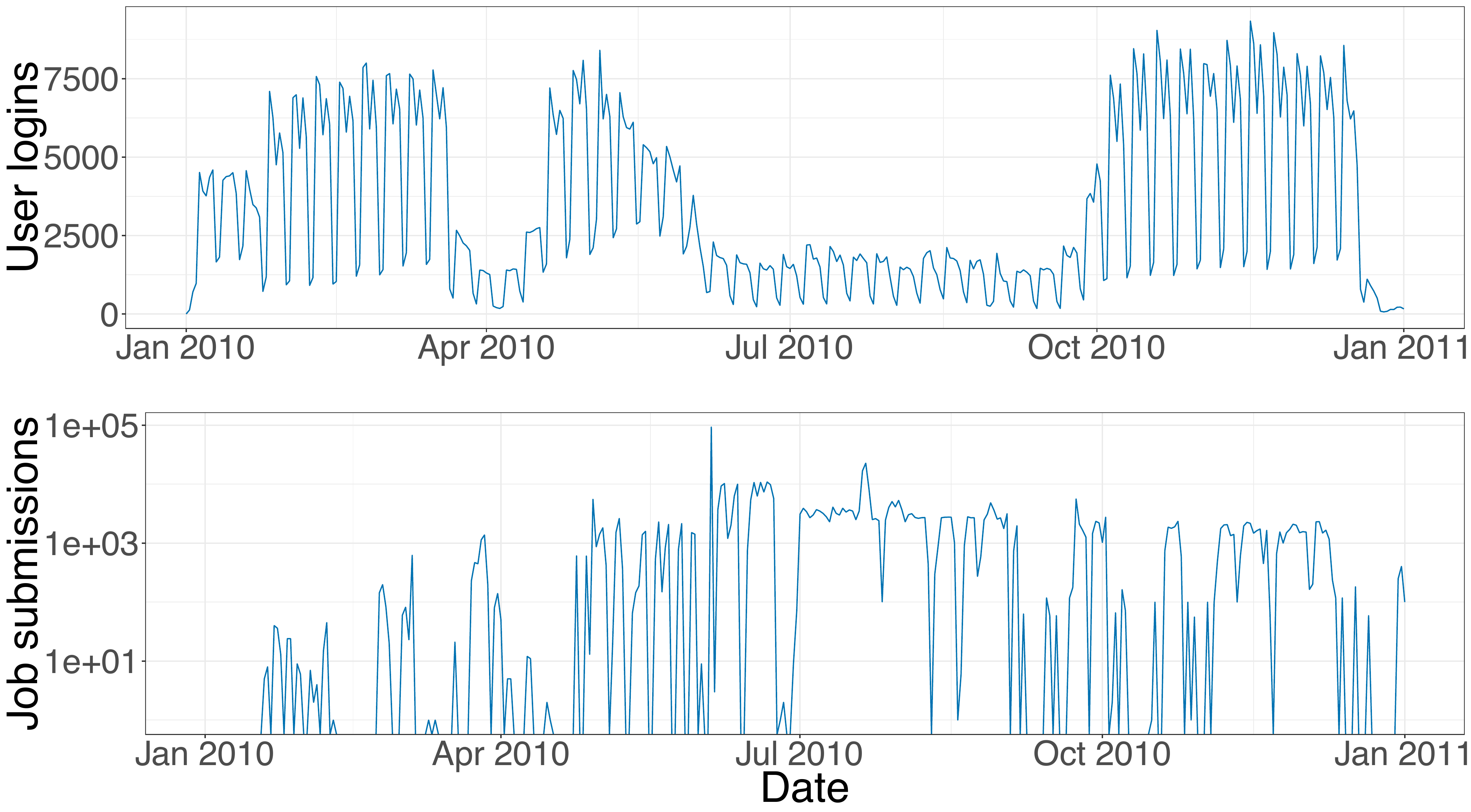}
\caption{Number of primary user logins per day and HTC tasks per day}
\label{fig:users}
\end{figure}

Each primary user record consists of a tuple of three elements: i) timestamp of user login, ii) name of computer, and, iii) timestamp of user logout. Each of the timestamps is to an accuracy of the nearest millisecond.

Analysis of the primary users (interactive users) reveals a strong seasonal influence to their usage patterns -- based around the construct of an academic year. For example Figure \ref{fig:users}-top illustrates the number of primary user logins per day. The three terms can be identified from the trace along with individual weeks.
Figure \ref{fig:users}-bottom, by contrast, shows the number of HTCondor tasks submitted per day. In this case there is no clearly identifiable pattern to the data. 

As a motivation for this work we produced a scheduler for the HTC-Sim system which broke the temporal rule for a simulation by allowing the scheduler (called Crystal) to know in advance the duration for each task along with the length of each idle period on each computer, thus providing a lower bound for the energy consumed by the system -- as no task will be terminated due to the computer returning to primary use. Table \ref{tab:crystal} provides the results along with the results for the default (random) scheduler. Indicating that we can save up to 73.6\% of the HTC energy. It may seem counterintuitive that the overhead has increased. This is a consequence of the scheduler which favours running longer tasks as opposed to the oldest task awaiting a computer. Hence, short-running tasks may see significantly more delay than expected.
\begin{table}[!b]
\renewcommand{\arraystretch}{1.3}
\caption{Potential energy saving from knowing task length and idle duration}
\label{tab:crystal}
\centering
\begin{tabular}{|l|r|r|}
\hline
{\bf Scheduler} & {\bf Overhead (mins)} & {\bf Energy (MWh)}\\
\hline
Random & 14.6140 & 121.5287 \\\hline
Crystal & 20.8989 & 32.0741 \\
\hline
\end{tabular}
\end{table} 

\section{Related Work}
\label{related}

Previous work by McGough {\em et al.}~\cite{suscom-short}, using HTC-Sim, has used Reinforcement Learning to identify computers less likely to be needed for primary use. Achieving an energy reduction of up to 53\%. However, this was only possible by significantly increasing the overhead. Instead we apply an alternative machine learning approach which allows us to predict the idle time for each computer -- using these for task execution. This allows us to reduce the energy consumed by up to 51\% without significantly increasing the overheads.

Machine learning is seeing increased use in optimising the operation of High Throughput and High Performance Computing environments. We do not seek to provide an exhaustive survey here, rather to highlight prominent works and the various areas of opportunity for machine learning to improve the performance of HTC systems.

\subsection{Scheduling Decisions:} The scheduling decisions made within HTC and HPC environments are typically governed by heuristics, taking as input characteristics of the workload and available computational resources. The use of machine learning opens up the opportunity to move from fixed heuristic to dynamic policies for scheduling workloads. Carastan-Santos~\emph{et al}~\cite{carastan2017obtaining} provide one such work, demonstrating notable improvements in task slowdown against existing scheduling approaches.

\subsection{Resource Allocation:} A substantial body of work has focused on the resource selection and allocation using machine learning. In particular, Reinforcement Learning (RL) has been shown to provide significant improvements over naive approaches. Bod\'{\i}k {\em et al.}~\cite{mlDC-short} apply RL for a datacenter workload subject to QoS constraints. Galstyan \emph{et al.}~\cite{galstyan2004resource} applied \emph{Q}-learning with an $\varepsilon$-greedy selection rule for resource selection in a grid environment. Tesauro \emph{et al.}~\cite{tesauro2006hybrid} proposed the Sarsa(0) approach for resource allocation in multiple server hosting environments for web applications. Several works specifically target energy conservation using RL, e.g. Das \emph{et al.}~\cite{das2008power}.

\subsection{Runtime prediction:} Accurate estimates of task execution times can be used to improve scheduler decision making. However, users of clusters have been shown to provide poor estimates of task runtimes~\cite{lee}. Empirical studies from production environments have shown tasks typically to consume only 30\% of the estimated time~\cite{cirne, ward, chiang}. This sometimes arises due to premature termination due to misconfiguration~\cite{alern}, due to performance variability within often-heterogeneous clusters~\cite{nitzberg}, or due to mis-reporting by users to avoid the preemption of tasks which exceed their estimated runtime. Machine learning approaches have shown promise in tackling some of these challenges. For example, Gaussier~\emph{et al}~\cite{gaussier2015improving} adopt online regression to predict task running times, and in turn optimise backfilling strategies with respect to a task slowdown metric.

\subsection{User assistance:} Machine learning approaches have also been shown to be useful in assisting the users of HPC systems. Rodrigues~\emph{et al}~\cite{rodrigues2016helping} demonstrate a tool which uses multiple model types (including Multilayer Perceptron and Random Forest, as we use in this work) to assist users to predict the memory requirements for their workloads. The authors found no single model provided the best predictions, so adopt an ensemble of several models. The accurate prediction of memory utilisation opens up the potential for more efficient resource allocation and workload consolidation.

\section{Prediction of computer idle time through Machine Learning}
\label{ml}
We describe here the process of converting the interactive user trace-log into a format which can be used for the purposes of machine learning along with the two machine learning algorithms used. We train a model for each computer individually as this gave slightly lower Mean Squared Error (MSE) than working at the cluster or whole system level -- more significantly the training time was much shorter. Numerous machine learning algorithms were evaluated, however, as the data was highly non-linear, Random Forest and MultiLayer Perceptron produced the best results (lowest MSE). We predict a months worth of interactive user data based on a number of previous months worth of data.

It should be noted that we do not consider predicting the duration of tasks -- although it is possible to train a machine learning algorithm accurately on a set of tasks~\cite{McGough:2017:UML:3053600.3053612-short}, it is not easy to predict future tasks durations. As although there is significant local correlation between tasks, this correlation dissipates quickly -- e.g. a user may submit thousands of tasks of similar duration at the same time, however, once these are complete their future work differs markedly.

\begin{figure}[]
\centering
\includegraphics[width=8cm]{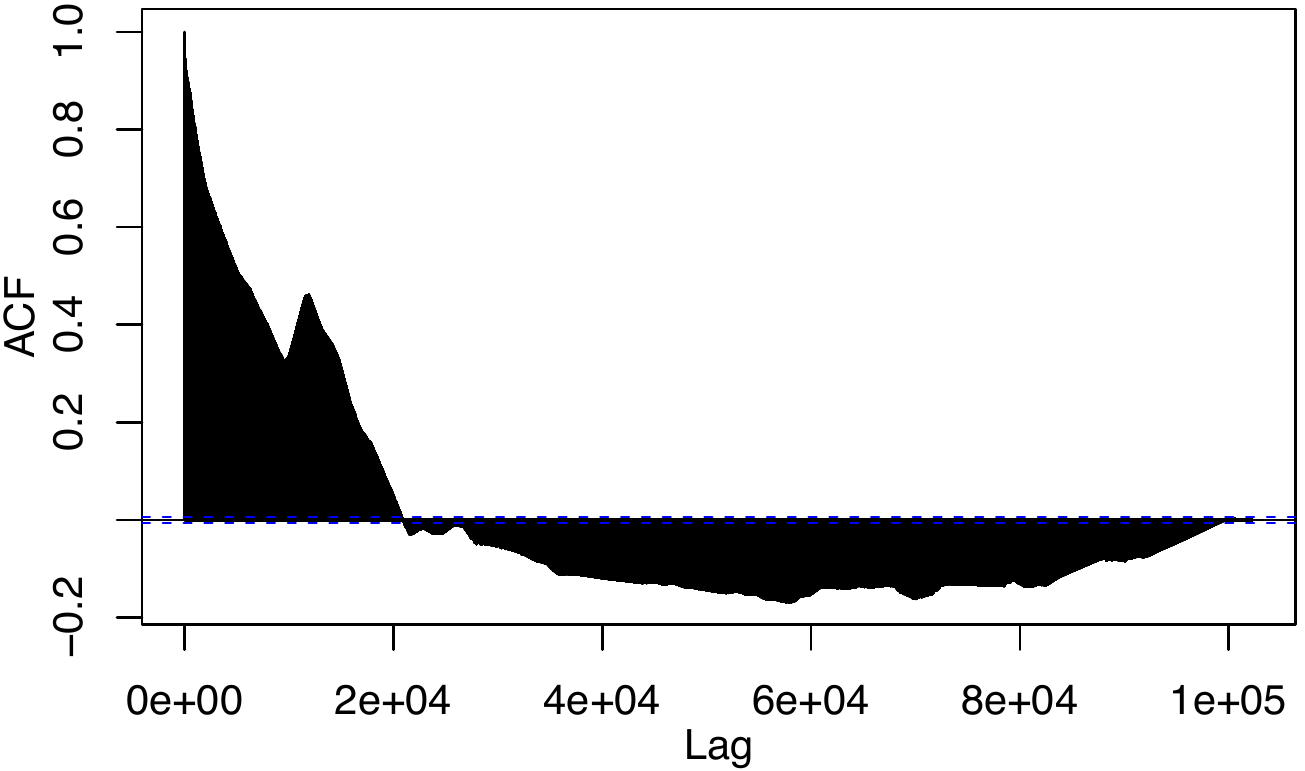}
\caption{Auto Correlation Function for task duration}
\label{fig:lag}
\end{figure}

Figure \ref{fig:lag} presents an autocorrelation plot (correlogram) for task durations within the trace. The plot presents levels of autocorrelation for data values at varying time lags. Values deemed significantly non-zero fall outside of the blue dotted lines. We see a very strong positive correlation for lags up to 20000, demonstrating that successive tasks are likely to exhibit similar durations. For lag values greater than 20000 we see negative correlation, arising from the interplay between dominant groups of tasks exhibiting relatively large and relatively small durations.
\subsection{Data}
\begin{figure}[!b]
\centering
\includegraphics[width=6.51cm]{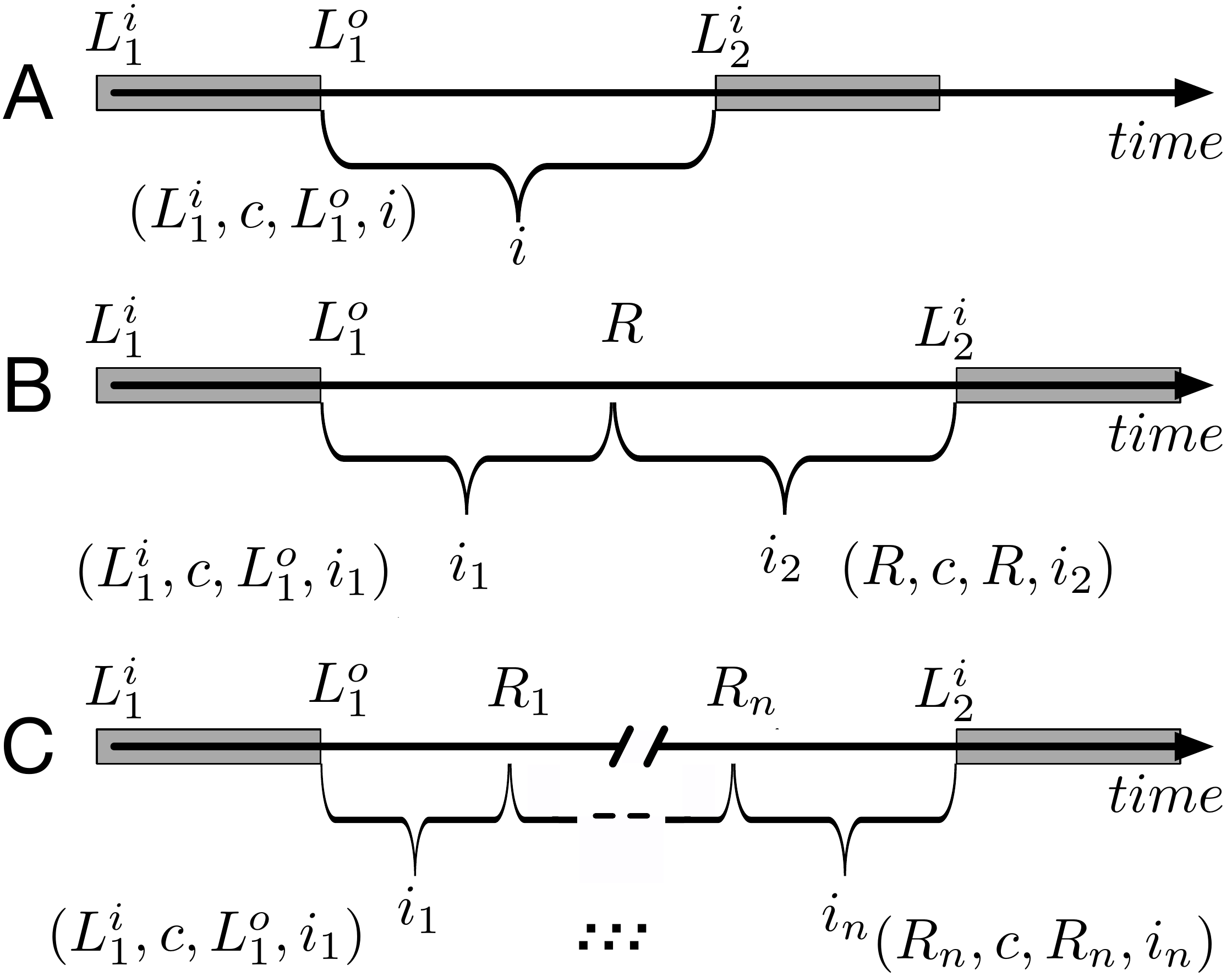}
\caption{Timeline for computer idle periods}
\label{fig:time}
\end{figure}
Each interactive user interaction is a tuple:
$$
(L^i, c, L^o)
$$
where $L^i$, $L^o$ are the login and logout times, and $c$ is the computer used. We extend this with a predicted idle time:
$$
(L^i, c, L^o, i)
$$
where $i$ is the actual or predicted time between two consecutive interactions on the same computer.
\subsubsection{Dealing with Reboots}
There are two types of event which cause a task to terminate: logging in of an interactive user, or, an automated computer reboot -- for software updates and system clearing. If an interactive user is logged in at reboot time then the reboot is postponed until the user logs out. This effectively extends the interactive user's login time and can be ignored here. However, if the reboot happens during an idle period then the idle period is split into multiple shorter idle periods. Figure \ref{fig:time} illustrates the different scenarios. In case A there is no reboot between a logout and the next login, therefore the idle period is equivalent to the time difference between the logout and login events. Case B contains a reboot between the logout and the next login event. We place a `synthetic' interactive user at the time of the reboot -- splitting the idle interval. Note that these synthetic interactive users consume zero time and once we have predicted the idle periods they are extracted into a separate file used for setting the predicted idle time after reboots. Case C is the extension of Case B where there are no interactive user login events over several days. Thus multiple `synthetic' interactive users are entered for each reboot time. As these computers have very little interactive user activity these are potentially the best computers to run HTC tasks on and hence we wish to capture this information.
\subsubsection{Dealing with sparse data}
Sparsity of the interactive user log when used for training can reduce the ability for the machine learning approaches to accurately predict -- especially for points far from the training data -- e.g. a computer in a locked cluster room will only have synthetic reboots in the log and will lack training data for what may happen far from the reboots. The same applies if there are long periods between interactive users. To overcome this we create a further set of `synthetic' interactive users at regular intervals between each logout and subsequent login (including reboot synthetic users). For a logout at time $L^o$ and subsequent login at time $L^i$ we create the set of `synthetic' (overlapping) users:
\begin{eqnarray}
\{(L^i, c, L^o, i), (L^i + \delta, c ,L^o, i - \delta), ..., \nonumber \\
 \{(L^i + n\delta, c, L^o, i - n\delta)\},\nonumber
\end{eqnarray}
such that $n = \lfloor i / \delta \rfloor$, and $\delta$ is the time between these synthetic interactive user records. This significantly improves the ability of the machine learning approaches to predict. 
\subsubsection{Preparation of data for Machine Learning}
Machine learning algorithms require numeric data ideally in the range of -1 to 1. As our trace-log contains timestamps and computer names these need converting:

{\bf Timestamps:} Can be converted into a Unix epoch. However, this does not make such concepts as day of the week, month or term available as features. Thus we add to the tuple the following features extracted from the logout time:
\begin{itemize}
	\item {\bf activeDuration:} duration for the last interactive user,
	\item {\bf minute:} number of minutes past the hour,
	\item {\bf hourOfDay:} hour of the day,
	\item {\bf dayOfWeek:} day of the week,
	\item {\bf dayOfMonth:} day of the month,
	\item {\bf month:} month of the year,
	\item {\bf year:} the year,
	\item {\bf termWeeks:} week of the current term -- in the range 1 -- 10, with -1 representing outside of the term,
	\item {\bf term:} term [1--3] with -1 indicating outside of a term.
\end{itemize}

{\bf Computer:} One hot encoding could be used where each computer becomes a separate feature. However, as we have 1386 computers this would lead to an equivalent number of features, slowing down training and prediction. Instead we use an integer encoder based on the cluster name, and the computer within that cluster -- only adding two features.

In order to scale our features to be in the range 0 -- 1 we divide each feature by the maximum value that the feature can take -- e.g. the dayOfMonth value can be divided by 31.

We now train the machine learning algorithms on several months' worth of training data (including actual idle times). Once trained we can predict the idle times for the following month (which lacks the idle times). As the data we are predicting on only contains information that would be known at the time the interactive user logs out we do not break the temporal rule for our simulation.

\subsection{Random Forest}
Random Forest~\cite{liaw2002classification} (RF) is a machine learning method which combines multiple decision trees into a single model and can be used for classification or regression. With regression being used in the work presented here. A RF creates a collection of decision trees each trained on a subset of the features. A decision tree is a construct where at each branching point a question is answered which moves one closer to the leaf containing the `correct' result. When predicting, each decision tree produces a predicted value with the modal value returned for the whole RF. An advantage of RF is that it is tolerant to overfitting to the data.

\begin{figure}[!t]
\centering
\includegraphics[width=8 cm]{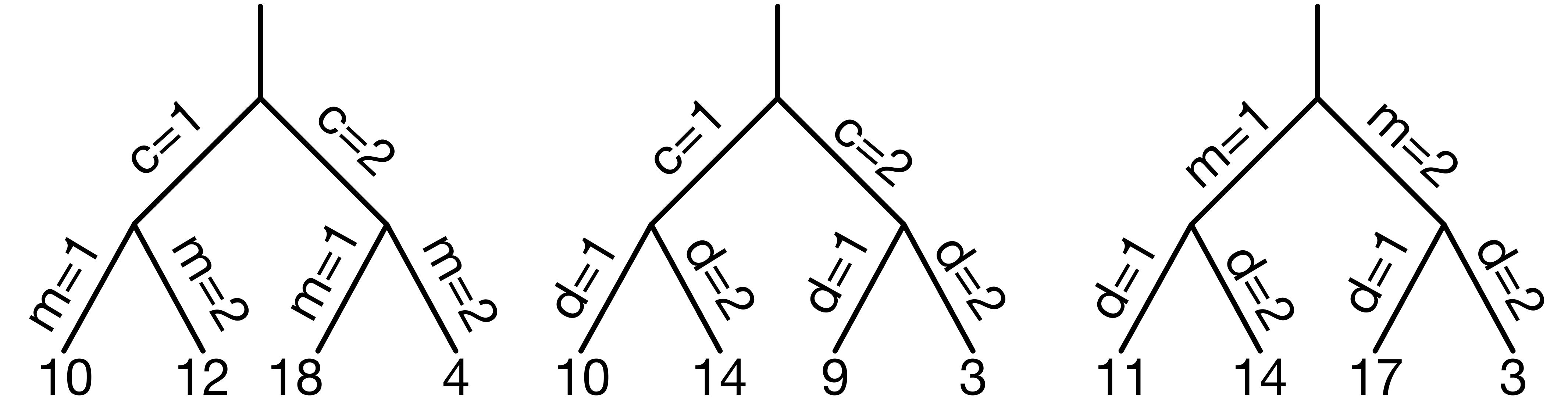}
\caption{A simple Random Forest}
\label{fig:random}

\end{figure}

Figure \ref{fig:random} depicts a simple RF of three trees derived from three features (c - cluster number, m - machine number, d - day of week) to predict the idle time. To predict the idle time for (c=1, m=1, d=1) each tree evaluates a prediction (10,10,11) with the value 10 returned.

\subsection{MultiLayer Perceptron}
A MultiLayer Perceptron~\cite{ruck1990multilayer} (MLP) is a feed-forward artificial neural network~\cite{DLBook} with a minimum of three fully connected layers -- an input layer, one or more hidden layers and an output layer. Each hidden and output layer contains a non-liner activation function, examples of which include ReLU, sigmoid or tanh. The MLP is trained through backpropagation, most commonly via some form of the Gradient Decent algorithm. 

The MLP works by taking a set of features in on the input layer. These values are fed forward to the next layer where they are scaled via a learnt weighting value at each node in the hidden layer, before being summed together. This summed value is then passed through the activation function before being fed forwards to the next layer. The network is trained by repeatedly feeding the network with input data and an output target label, where the prediction of the network is compared with the target label using a loss function -- mean squared error for regression tasks. The weights of the network are then tuned via backpropagation to minimise the error given by the loss function. 
\subsection{Ensemble Approaches}
A machine learning ensemble approach combines the predictions from multiple independently trained machine learning models to produce a `better' overall result. We present a number of ensemble techniques used to improve the accuracy of our predictions:
\begin{itemize}
	\item {\bf Max:} of the RF and MLP predictions. The expectation is that this will lead to a more speculative scheduler.
	\item {\bf Min:} of the RF and MLP predictions. The expectation is that this will lead to a more conservative scheduler.
	\item {\bf Average:} of the RF and MLP predictions. The expectation is that this will reduce extreme values.
	\item {\bf Last Month:} At the end of a month evaluate the MSE between the real and predicted values for both RF and MLP and use the lower for the following month. The expectation is that a good approach for one month is likely to be good for the next month.
	\item {\bf Best on average:} Extending the `Last Month' to take into account the best from all previous months.
\end{itemize}
\subsection{Scheduler for HTC-Sim}
Here we describe Machine Learning scheduler developed to work within HTC-Sim -- Algorithm \ref{alg:sched}. The algorithm first determines if the task has been attempted before and uses the longest attempt as a lower bound on the execution length. It then finds the idle computer with the longest predicted idle time and returns it (unless the predicted idle duration is less than the previous run attempt). If no suitable idle computer is free it attempts to find a sleeping computer which has at least the previous runtime free. If no sleeping computer is found then no computer is returned. Where runBefore is a boolean indicating if the task has been run previously, maxPreviousRunDuration returns the maximum duration of the previous run attempts, idle is the set of all currently idle computers, sleep is the set of all currently sleeping computers, findLongestIdle returns the computer from the set with the longest predicted idle time and predictedIdle computes the remaining idle time at time $t$ as the computer holds the idle time since the last logout. Hence:
$$
predictedIdle = lastLogout + idle - t.
$$
\subsection{Metrics}
We define task overhead as the time a task is within the HTC system and the time the task would take on a dedicated computer. The average overhead can be defined as:
$$
\frac{1}{|T|}\sum\nolimits^{t \in T} (f_{t} - s_{t} - d_{t})
$$
where $T$ is the set of tasks, $s_t$, $f_t$ are the submission and finish times of task $t$, and $d_t$ is the execution time of task $t$.
\begin{algorithm}[!t]
\caption{ML Scheduler}
\label{alg:sched}
\begin{algorithmic}[1]
\\MLScheduler(time t, task $\tau$) returns computer to use
\If{$\tau$.runBefore}
\State p = $\tau$.maxPreviousRunDuration
\Else
\State p = 0
\EndIf
\State c = idle.findLongestIdle
\If{c.predictedIdle(t) $>$ p}
\State return c
\EndIf
\State c = sleep.findLongestIdle
\If{c.predictedIdle(t) $>$ p}
\State return c
\EndIf
\State return null
\end{algorithmic}
\end{algorithm}
We define the energy consumption ($E_t$) for each task $t$:
$$
 E_t = \sum\nolimits^{k \in A_t}  (e_{t,k} - b_{t,k}) \cdot E_{t,k}
$$
where $A_t$ is the set of all attempts at task $t$, $E_{t,k}$ is the energy consumption rate of the computer chosen for attempt $k$ of task $t$, $e_{t,k}$ is the end time of attempt $k$ of task $t$, and $b_{t,k}$ is the corresponding start time. We can then compute the total energy consumed by summing $E_t$ for all $t$.
\section{Experimental Setup}
\label{experiment}
As HTC-Sim consumes a trace-log of interactive user and HTC tasks we have augmented the interactive user trace-log with the predicted idle values. We used Scikit-Learn (version 0.19.1) to generate the predicted RF and MLP values. 
\subsection{Data}
We run our experiments against the 2010 exemplar datasets used with HTC-Sim. We train the interactive user time in intervals of one month and train on all data from the start of 2009 until immediately before the start of the month to be predicted. I.e. to predict February 2010 we train on January 2009 through to January 2010 inclusive. 
\subsection{Parameters and Features}
\subsubsection{Maximising prediction accuracy through `synthetic' tasks} We performed a search-space analysis on $\delta$ -- the time interval between `synthetic' users when reducing data sparsity. Evaluating $\delta$ between 5 and 60minutes, achieved maximal accuracy at $\delta = 10$minutes. 
\subsubsection{Identification of optimal feature set} Although we could use all of our features as defined in Section \ref{ml} it is often the case that training on a subset will give better accuracy. RF allows for the identification of the most important features. This identified the best feature set as: $\{$epochLogin, epochLogout, activeDuration, hourOfDay, dayOfWeek, dayOfMonth, month$\}$.
\subsubsection{MLP hidden layers} The number of hidden layers and number of nodes per layer can significantly affect the accuracy of a MLP. We performed a search space analysis of all possible networks with up to four hidden layers and forty nodes per layer. This identified a four hidden layer MLP with 18, 14, 9, and 10 nodes per layer minimised the MSE for the majority of computers. The best activation function was RELU along with the Adam solver.
\section{Results}
\label{results}
\begin{table}[b]
\center
\caption{Energy and overhead of different approaches}
\label{tab:results}
\begin{tabular}{|l|r|r|r|r|}\hline
{\bf Scheduler}        & {\bf Overhead} & {\bf Energy} & {\bf Productive} & {\bf Wasted} \\\hline
                & (mins)   & (MWhs) & (MWhs) & (MWhs) \\\hline
Random          & 14.61    & 121.53       & 33.85      & 87.68\\\hline
Crystal         & 20.90    &  32.07       & 32.07      & 0 \\\hline
RF              & 12.81    &  60.95       & 33.96      & 26.99 \\\hline
MLP 	        & 12.94    &  63.72	      & 33.17      & 30.55 \\\hline
MAX	            & {\bf 11.80}	   &  63.51 	      & 32.89      & 30.63 \\\hline
MIN	            & 15.32    &  {\bf 59.12}	      & 33.91      & {\bf 25.21} \\\hline
Average         & 12.77    &  63.14       & 33.36      & 29.79 \\\hline
Last month      & 12.10    &  64.75       & 32.99      & 31.76 \\\hline
Best on average & 12.24    &  66.06       & {\bf 32.70}      & 33.36 \\\hline
\end{tabular}
\end{table}
Table \ref{tab:results} presents the overheads observed and energy consumed for all of the approaches presented within the paper. Random is the original task scheduler provided with HTC-Sim whist Crystal is the scheduler which breaks the temporal rules and has full knowledge of future events. Crystal is not considered here as a valid approach -- presented here only for comparison.  The lowest overhead is observed for the MAX ensemble approach -- which matches in with our assumption that this would be more speculative on deploying tasks to computers. It would appear that this approach paid off with tasks completing within the predicted idle time. Likewise MIN leads to the largest overhead for the machine learning approaches as it is more conservative when considering computers which may not have enough time to complete the task. This would suggest that the time waiting to find a computer to run a task has a more significant impact on overhead than the time incurred through aborted executions. All other machine learning approaches have overheads lower than the Random scheduler, though there is no significant difference between them. 

The energy consumption due to the HTC system is decomposed into productive energy consumed -- i.e. the successful execution of the task, and wasted HTC energy -- aborted executions. As the Crystal approach is fully aware of the future state of the system it has no wasted energy. It should be noted that the productive energy in all cases is approximately the same -- around 33MWh. The variation here is a consequence of different computers within the exemplar setup having different energy consumption rates -- different computers leads to different energy consumption. 

In all machine learning cases the energy consumption was brought down to a very similar value -- in the range 59.1 to 66.1MWh. The lowest energy consumption was observed for MIN (59.1MWh) which matches in with the assumption that this approach is being more conservative with computer selection. Though as noted above this leads to a slightly higher overhead. However, all machine learning approaches significantly reduce the energy consumed.

\begin{figure}[t]
\centering
\includegraphics[width=7cm]{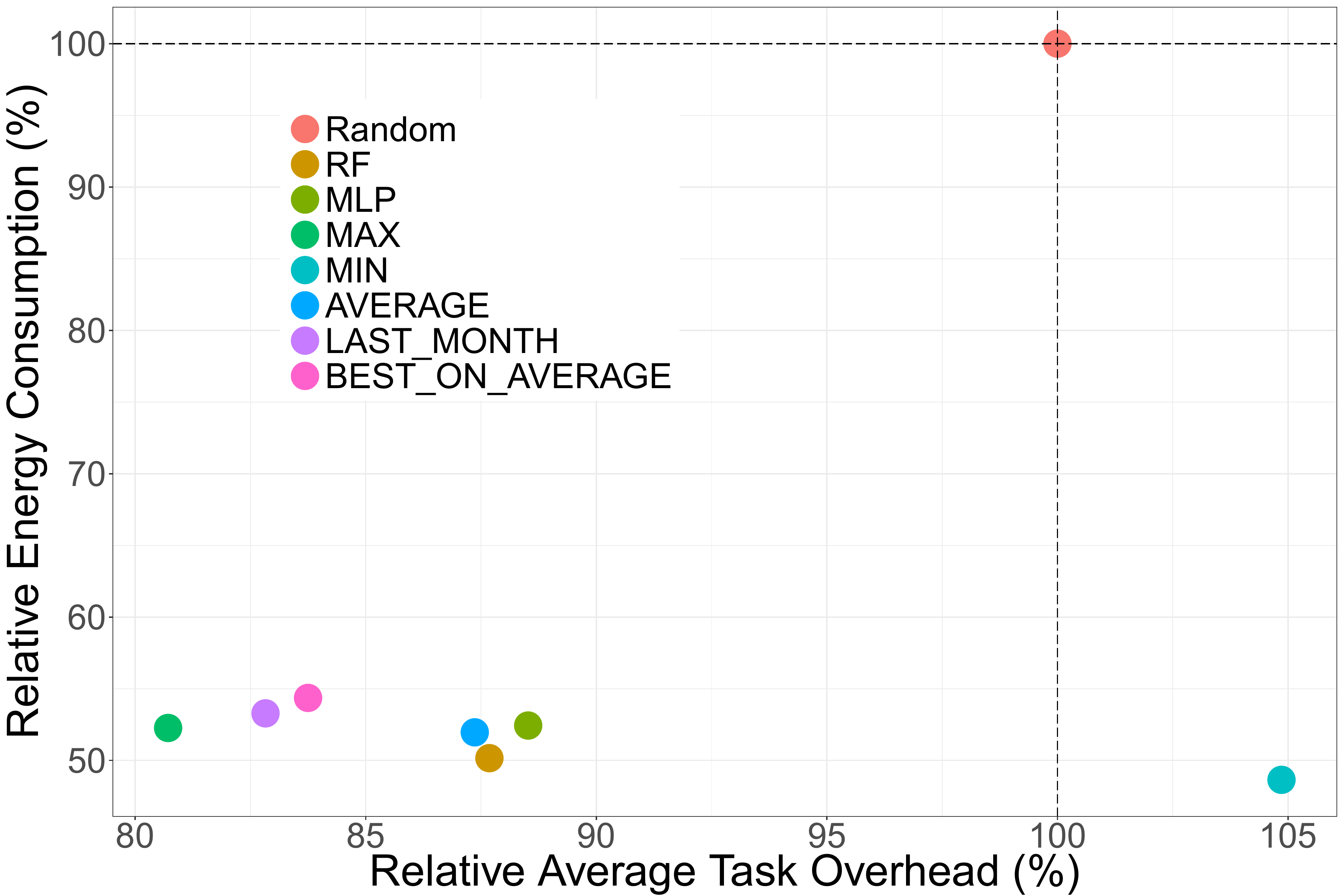}
\caption{Overhead vs. Energy for different machine learning approaches}
\label{fig:scatter}
\end{figure}
Figure \ref{fig:scatter} compares the energy to overhead for the different machine learning approaches. The values have been scaled to a percentage of the Random scheduler. Note that the Crystal scheduler is dropped here as it would not be a practical scheduler in the real world. The graph is broken up into four quadrants, by the dotted line, where the bottom left quadrant beats the Random scheduler in both energy and overhead. All machine learning schedulers except for MIN fall into this quadrant. Although the MIN approach fails to beat the overhead it does have the lowest energy consumption rate and only increases the overhead by 4.9\%. Hence, if energy saving is the most important consideration then MIN would be best -- saving 51.4\% energy. Alternatively if overhead is more important then Max would reduce energy consumption by 47.7\% and overhead by 19.3\%. All other approaches provide similar performance to MAX.

Table \ref{tab:trsin} illustrates the training time (in seconds) for the two machine learning approaches. Both exhibit a linear relationship with the number of months of training data, with MLP scaling better when the number of months increase. This is not considered a significant impact on the task execution time and can be computed off-line. Prediction impact is negligible taking less than 5ms for RF and 0.03ms for MLP and is independent of the number of months.

\begin{figure}[b]
\centering
\includegraphics[width=8 cm]{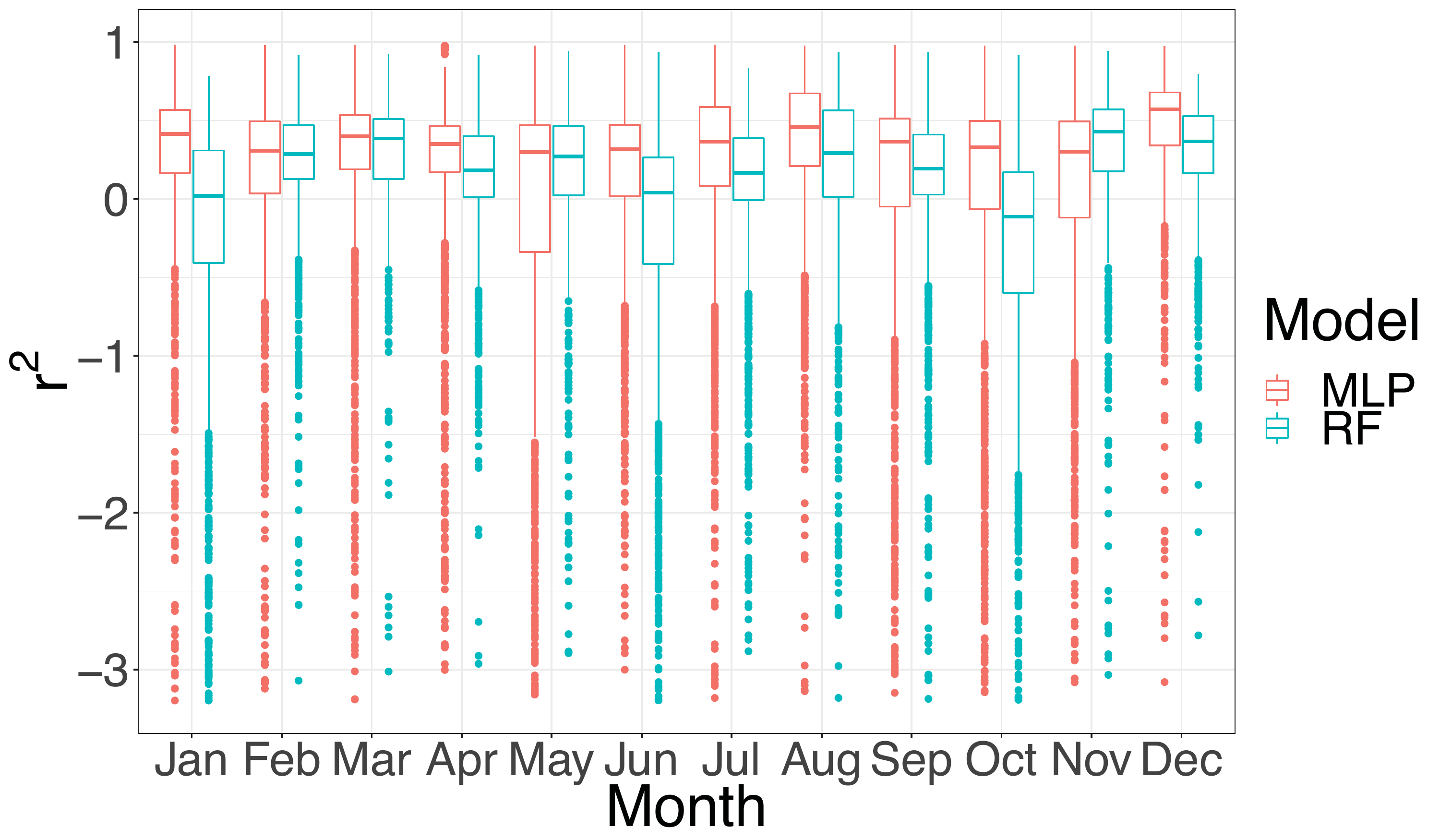}
\caption{Accuracy vs. months of training}
\label{fig:accuracy}
\end{figure}

Figure \ref{fig:accuracy} presents a box and whisker plot for the $r^2$ value for each computer within the cluster for each month. The $r^2$ value is a statistical representation of how well the predicted idle time matches with the real idle time. A value of 1 indicating a perfect match with lower values indicating a less accurate match -- with no lower bound on `badness'. In all cases but one the median value is greater than zero -- where zero implies a constant value would be as good. The only exception to this is RF for October -- which could be a consequence of the fact that RF is thrown by the start of a new term. In all cases MLP has a median value closer to one than RF, which would suggest that it should work better for the simulation. As this is not the case it would suggest that although MLP is more accurate it predicts values which are too large more often than RF. It should be noted that Figure \ref{fig:accuracy} has been clipped at -3.2. Although there are some outlier points below this value these represent only a small fraction of the 1386 values. 
\begin{table}\center
\caption{Energy and overhead of different approaches}
\label{tab:trsin}
\setlength{\tabcolsep}{4.2pt}
\begin{tabular}{|l|r|r|r|r|r|r|r|r|r|r|r|r|r|}\hline
{\bf App}        & \multicolumn{12}{c|}{\bf Months of Training} \\\hline
    & 12  & 13  & 14  & 15  & 16  & 17  & 18  & 19  & 20  & 21  & 22  & 23 \\ \hline
RF  & 3.5 & 3.8 & 4.1 & 4.4 & 4.7 & 5.1 & 5.4 & 5.8 & 6.1 & 6.5 & 7.1 & 7.3 \\\hline
MLP & 3.8 & 4.0 & 4.1 & 4.2 & 4.5 & 4.6 & 4.7 & 4.8 & 5.0 & 5.2 & 5.4 & 5.4 \\\hline
\end{tabular}
\end{table}
%
\section{Conclusion}
\label{conc}
In this paper we have utilised two machine learning algorithms -- Random Forest (RF) and MultiLayer Perceptron (MLP) -- to predict the amount of idle time between consecutive primary user activity on a computer within a volunteer computing environment. Allowing us to develop a scheduler which targets work to those computers with the longest expected idle time. Reducing the amount of energy consumed by reducing the number of aborted task executions due to the primary user wishing to make use of their computer. To improve the results we use ensemble approaches to combine the different machine learning algorithms.

We demonstrate, through the use of simulation that we can save between 45.6\% and 51.4\% of the energy consumed through the High Throughput Computing (HTC) system with minimal impact on the overhead observed by the HTC user -- in the worst case increasing the overhead by 4.9\%. If energy saving is the primary goal then taking the minimum of the RF and MLP predictions is best -- saving 51.4\%. However, this increases the overhead by 4.9\%. If reducing the overhead is most important then taking the maximum of RF and MLP will reduce the overhead by 19.3\% whilst still decreasing the energy consumption by 47.7\% -- only 3.6MWh more per year.

We re-train the machine learning algorithms on a monthly basis. However, it may be possible to improve energy savings by performing re-training more frequently. Currently we take no account of the execution time of the tasks -- save for using prior (aborted) runs to provide a minimum -- although this has proven to be a hard value to predict {\em a priori} it may be possible to provide course intervals and classify tasks into these. In addition, as there is strong local correlation between tasks submitted at the same time,  it may be possible to exploit this local correlation by predicting task duration based on other temporally close tasks.

Although our approach uses real trace-logs, allowing for complex situations to occur, it would be good to deploy this work into a real HTC environment to evaluate it in real-time.
\let\oldthebibliography\thebibliography
\let\endoldthebibliography\endthebibliography
\renewenvironment{thebibliography}[1]{
  \begin{oldthebibliography}{#1}
    \setlength{\itemsep}{0.06em}
    \setlength{\parskip}{0.07em}
}
{
  \end{oldthebibliography}
}
\bibliographystyle{IEEEtran}
\bibliography{IEEEabrv,../../bib/HTC,../../bib/US,../../bib/Scheduling,../../bib/Green,../../bib/Checkpointing,../../bib/AI,../../bib/Cloud,../../bib/sim}
\end{document}